\documentclass[12pt]{article}
\usepackage{amsmath,amssymb}
\usepackage{graphicx}
\newcommand{\eqdef}{\stackrel{\text{def}}{=}}

\newcommand{\n}{\nonumber \\}
\newcommand{\bm}{\boldsymbol}
\newcommand{\ignore}[1]{}

\numberwithin{equation}{section}
\newcommand{\Romannumeral}[1]{\uppercase\expandafter{\romannumeral#1}}

\newcommand{\ma}{\hspace{0pt}}
\newcommand{\cX}{\mathcal{X}}

\allowdisplaybreaks[4]

\begin{document}

\baselineskip=20pt
\newcommand{\preprint}{
\vspace*{-20mm}\begin{flushleft}\end{flushleft}
}
\newcommand{\Title}[1]{{\baselineskip=26pt
  \begin{center} \Large \bf #1 \\ \ \\ \end{center}}}
\newcommand{\Author}{\begin{center}
  \large \bf 
  Ryu Sasaki${}$ \end{center}}
\newcommand{\Address}{\begin{center}
     Department of Physics and Astronomy, Tokyo University of Science,
     Noda 278-8510, Japan
        \end{center}}
\newcommand{\Accepted}[1]{\begin{center}
  {\large \sf #1}\\ \vspace{1mm}{\small \sf Accepted for Publication}
  \end{center}}

\preprint
\thispagestyle{empty}

\Title{Exactly solvable multicomponent spinless fermions}

\Author

\Address
\vspace{1cm}

\begin{abstract}
 By generalising the one to one correspondence between exactly solvable hermitian
 matrices $\mathcal{H}=\mathcal{H}^\dagger$ and exactly solvable spinless 
 fermion systems $\mathcal{H}_f=\sum_{x,y}c_x^\dagger\mathcal{H}(x,y)c_y$,
 four types of exactly solvable multicomponent fermion systems are constructed explicitly.
 They are related to the multivariate Krawtcouk, Meixner and 
 two types of Rahman like polynomials, constructed recently by myself.
 The Krawtchouk and Meixner polynomials are the eigenvectors of certain real symmetric
 matrices $\mathcal{H}$ which are related to the difference equations governing them.
 The corresponding fermions have nearest neighbour interactions.
 The Rahman like polynomials are eigenvectors of certain reversible Markov chain matrices $\mathcal{K}$,
 from which real symmetric matrices $\mathcal{H}$ are uniquely defined by the similarity transformation
 in terms of the square root of the stationary distribution. The fermions have wide range interactions.
 \end{abstract}

%
%
\section{Introduction}
\label{sec:intro}

Orthogonal polynomials have played important roles in mathematics, 
physics and other disciplines in science and technology.
Among them, the hypergeometric orthogonal polynomials of 
Askey scheme \cite{askey}--\cite{os13}
occupy the center stage, as they satisfy differential or difference equations 
on top of the three term recurrence relations.
Recently two multivariate hypergeometric orthogonal polynomials of Aomoto-Gelfand \cite{AK,gelfand}
are constructed explicitly by myself \cite{mKra}. 
They  are multivariate Krawtchouk and Meixner polynomials satisfying multivariate difference equations
which are direct generalisation of those for the single variable Krawtchouk and Meixner polynomials.
They constitute the eigenvectors of real symmetric matrices (Hamiltonians) 
obtained from the difference equations in terms of similarity transformations.
Two types of Rahman like polynomials   are also constructed as another 
Aomoto-Gelfand type hypergeometric orthogonal polynomials \cite{Rah}. 
They are eigenvectors of certain reversible Markov chain matrices \cite{gr1}--\cite{gr3},
which are equivalent to 
real symmetric matrices (Hamiltonians) by similarity transformations in terms of the
stationary distribution.

As possible applications in physics of these new objects, four multivariate hypergeometric orthogonal polynomials,
I show that their Hamiltonians define  four {\em exactly solvable multicomponent spinless fermion systems}.
The construction method is the simple multidimensional generalisation of the one dimensional 
exactly solvable lattice fermion systems \cite{solvfermi,widefermi}. 
For  different constructions of multi-component inhomogeneous free fermions, see \cite{bernard}.

The interesting and important theories of multivariate Krawtchouk, Meixner and 
Rahman polynomials have been developed by many authors
over  a long period; Griffiths \cite{Gri1}--\cite{Gri2}, Cooper-Hoare-Rahman \cite{coo-hoa-rah77,HR0},
Tratnik \cite{tra}, Zhedanov \cite{zheda}, Mizukawa \cite{mizu1,mizu}, Mizukawa-Tanaka \cite{mt},
Iliev-Xu \cite{ilxu}, Hoare-Rahman \cite{HR},
Gr\"unbaum \cite{gr1}, Gr\"unbaum-Rahman \cite{gr2,gr3}, Iliev-Terwilliger \cite{IT}, Iliev \cite{I11}--\cite{I23},
Genest-Vinet-Zhedanov \cite{genest}, Diaconis-Griffiths \cite{diaconis13}, Xu \cite{xu}, as explained 
appropriately in \cite{mKra,Rah}.
None of these preceding polynomials, however, are eigenpolynomials of hermitian matrices.
Their details are not relevant for the present purpose of constructing exactly solvable multicomponent
fermions.

This paper is organised as follows.
In section two, after a brief summary of the basic properties of the multivariate Krawtchouk polynomials
in \S\ref{sec:mKrapoly}, exactly solvable fermion systems corresponding to the multivariate Krawtchouk
polynomials are introduced in \S\ref{sec:mKraf}. 
The main results of the multivariate Meixner polynomials are recapitulated in \S\ref{sec:mMeipoly}.
In \S\ref{sec:mMeif}, the exactly solvable fermion system Hamiltonian on a multidimensional semi-infinite
integer lattice is presented based on the Hamiltonian of the multivariate Meixner polynomials 
derived in \S\ref{sec:mMeipoly}.
A r\'esum\'e of the common structure of Rahman like polynomials of type (1) and (2) 
are given at the beginning of section four.
The explicit forms of type (1) Rahman like polynomials, the Hamiltonian  and the corresponding exactly
solvable multicomponent fermion Hamiltonian with wide range interactions are shown in \S\ref{sec:Rahpoly1}.
The differences between type (1) and (2) Rahman like polynomials are pointed out in \S\ref{sec:Rahpoly2}.

Many explicit examples of exactly solvable theories demonstrated here  and in \cite{solvfermi,widefermi}
would offer a good laboratory to evaluate various interesting physical
quantities, {\em e.g.} entanglement entropy, etc \cite{bernard},\cite{gvz}--\cite{latorre}.
It would be crucial  to pick up the  effects of multidimensionality.
\section{Multi-Krawtchouk fermions }
\label{sec:Kraf}

Multivariate Krawtchouk polynomials, as the simplest example of multivariate hypergeometric
orthogonal polynomials, have been approached from many different angles, see for example
\cite{Gri1, Gri2, mizu,mt,I23,diaconis13}.
My approach  \cite{mKra} is rather different from them. 
Iliev \cite{I23} used similar method and obtained multivariate Krawtchouk polynomials 
depending on $2(n+1)$ parameters in contrast to $n$ parameters  in my approach.
Let us briefly summarise the main results of multivariate Krawtchouk polynomials \cite{mKra},
which will be cited as I.

\subsection{Multivariate Krawtchouk polynomials}
\label{sec:mKrapoly}

The  multivariate Krawtchouk polynomials $\{P_{\bm m}({\bm x})\}$ are 
 defined by two positive integers $N$ and $n$ ($N>n\ge2$) and $n$  distinct positive numbers
$p_i>0$,  $i=1,\ldots, n$, 
on a finite $n$-dimensional integer lattice $\mathcal{X}$,
\begin{equation}
 \bm{x}=(x_1,\ldots,x_n)\in\mathbb{N}_0^n,\quad  |x|\eqdef \sum_{i=1}^nx_i,\quad
 \mathcal{X}=\{\bm{x}\in\mathbb{N}_0^n\ |\, |x|\le N\}.
 \label{XKdef}
\end{equation}
The multivariate Krawtchouk $\{P_{\bm m}({\bm x})\}$ are orthogonal with respect to the 
{\em multinomial distribution}
 with probabilities $\{\eta_i\}$ which are functions of 
$\{p_i\}$,
\begin{align}
&\hspace{3cm} \sum_{{\bm x}\in\cX}W(\bm{x},N,\eta)P_{\bm m}({\bm x})P_{\bm {m}'}({\bm x})=0,
\quad \bm{m}\neq\bm{m}'\in\cX,
\label{pKortho}\\
&\hspace{2cm}W(\bm{x},N,\eta)\eqdef\frac{N!}{x_1!\cdots x_n!x_0!}\prod_{i=0}^n\eta_i^{x_i}
=\binom{N}{\bm{x}}\eta_0^{x_0}\bm{\eta}^{\bm{x}},
\label{WKr}\\
&\hspace{4cm}  x_0\eqdef
N-|x|,\  \binom{N}{\bm{x}}\eqdef\frac{N!}{x_1!\cdots x_n!x_0!},\n
&\quad \eta_i\eqdef\frac{p_i}{1+\sum_{j=1}^np_j},\quad \eta_0\eqdef\frac1{1+\sum_{i=1}^np_i},
\quad \sum_{i=0}^n\eta_i=1, \quad \bm{\eta}^{\bm{x}}\eqdef\prod_{i=1}^n\eta_i^{x_i}.
\label{etadef}
\end{align}

They form a complete set of eigenvectors of a real symmetric $|\cX|\times|\cX|$ matrix $\mathcal{H}$,
\begin{align}
&B_i(\bm{x})\eqdef N-|x|,\quad  D_i(\bm{x})\eqdef p_i^{-1}x_i,\qquad  i=1,\ldots,n,
\label{BDKrdef}\\
\mathcal{H}(\bm{x},\bm{y})&
\eqdef\sum_{j=1}^n\left[\bigl(B_j(\bm{x})+D_j(\bm{x})\bigr)\,\delta_{\bm{x}\,\bm{y}}
-\sqrt{B_j(\bm{x})D_j(\bm{x}+\bm{e}_j)}\,\delta_{\bm{x}+\bm{e}_j\,\bm{y}}\right.\n
&\left.\hspace{4cm}
-\sqrt{B_j(\bm{x}-\bm{e}_j)D_j(\bm{x})}\,\delta_{\bm{x}
-\bm{e}_j\,\bm{y}}\right],\qquad \qquad \bm{x},\bm{y}\in\cX,
\label{Hdef}\\
&\sum_{\bm{y}\in\cX}\mathcal{H}(\bm{x},\bm{y})\sqrt{W(\bm{y},N,\eta)}\,P_{\bm m}(\bm{y})
=\mathcal{E}(\bm{m})\sqrt{W(\bm{x},N,\eta)}\,P_{\bm m}(\bm{x}),\quad \bm{m}\in\cX,
\end{align}
in which $\bm{e}_j$ is the  $j$-th unit vector, $j=1,\ldots,n$. 
The eigenvalue $\mathcal{E}(\bm{m})$ has a linear spectrum
\begin{equation}
\mathcal{E}(\bm{m})\eqdef\sum_{j=1}^nm_j\lambda_j\ge0,\qquad \bm{m}\in\cX,
\tag{I.3.13}
\end{equation}
in which $\lambda_j>0$ is the $j$-th root of a degree $n$ characteristic polynomial $\mathcal{F}(\lambda)$
of an $n\times n$ positive definite symmetric matrix $F(p)$  depending on $\{p_i\}$,
\begin{equation}
0=\mathcal{F}(\lambda)\eqdef Det\bigl(\lambda I_n-F(p)\bigr),\quad F(p)_{i\,j}\eqdef 1+p_i^{-1}\delta_{i\,j}.
\tag{I.3.14}
\end{equation}
The multivariate Krawtchouk  polynomial $P_{\bm{m}}(\bm{x})$ is a terminating   $(n+1,2n+2)$ hypergeometric function 
of Aomoto-Gelfand  {\rm \cite{AK,gelfand,mizu}}
\begin{gather}
\label{PmK}
P_{\bm{m}}(\bm{x})
\eqdef \sum_{\substack{\sum_{i,j}c_{ij}\leq N\\
(c_{ij})\in M_{n}({\mathbb N_{0}})}}
\frac{\prod\limits_{i=1}^{n}(-x_{i})_{\sum\limits_{j=1}^{n}c_{ij}}
\prod\limits_{j=1}^{n}(-m_{j})_{\sum\limits_{i=1}^{n}c_{ij}}}
{(-N)_{\sum_{i,j}c_{ij}}} \; \frac{\prod(u_{ij})^{c_{ij}}}{\prod c_{ij}!},
\end{gather}
in which $M_{n}({\mathbb N}_{0})$ is the set of square matrices of degree $n$ with nonnegative integer
elements. 
Here, $(a)_n$ is the shifted factorial defined for $a\in\mathbb{C}$ and a nonnegative integer $n$,
$(a)_0=1$, $(a)_n=\prod_{k=0}^{n-1}(a+k)$, $n\ge1$.
The $n\times n$ matrix $u_{i\,j}$  is defined by
\begin{equation}
u_{i\,j}\eqdef\frac{\lambda_j}{\lambda_j-p_i^{-1}}=\frac1{1-p_i^{-1}\lambda_j^{-1}},\quad i,j=1,\ldots,n.
\tag{I.3.16}
\end{equation}
With the explicit expression of the multivariate Krawtchouk polynomials \eqref{PmK},  
the orthogonality relation \eqref{pKortho}
 now reads
\begin{align}
\sum_{\bm{x}\in\mathcal{X}}W(\bm{x},N,\eta)P_{\bm{m}}(\bm{x})P_{\bm{m}'}(\bm{x})
&=\frac{\delta_{\bm{m}\,\bm{m}'}}{\binom{N}{\bm{m}}(\bar{\bm{p}})^{\bm{m}}},\qquad 
(\bar{\bm{p}})^{\bm{m}}\eqdef\prod_{j=1}^n\bar{p}_j^{m_j},
\label{nKrorth}\\
\bar{p}_j&=\Bigl(\sum_{i=1}^n\eta_iu_{i,j}^2-1\Bigr)^{-1}>0,\quad j=1,\ldots,n,
\tag{I.3.18}
\end{align}
leading to the complete set of orthonormal eigenvectors $\{\hat{\phi}_{\bm m}(\bm{x})\}$,
\begin{align}
&\sum_{\bm{y}\in\mathcal{X}}\mathcal{H}(\bm{x},\bm{y})\hat{\phi}_{\bm{m}}(\bm{y})
=\mathcal{E}(\bm{m})\hat{\phi}_{\bm{m}}(\bm{x}),\quad \bm{x},\bm{m}\in\cX,
\label{mKHphin}\\
\sum_{\bm{x}\in\mathcal{X}}\hat{\phi}_{\bm{m}}(\bm{x})\hat{\phi}_{\bm{m}'}(\bm{x})
&=\delta_{\bm{m},\bm{m}'}, \quad
\sum_{\bm{m}\in\mathcal{X}}\hat{\phi}_{\bm{m}}(\bm{x})\hat{\phi}_{\bm{m}}(\bm{y})
=\delta_{\bm{x},\bm{y}}, \qquad \bm{x},\bm{y}, \bm{m},\bm{m}'\in\mathcal{X},
\label{ortrel}\\
\hat{\phi}_{\bm{m}}(\bm{x})&\eqdef \sqrt{W(\bm{x},N,\eta)}
P_{\bm{m}}(\bm{x})\sqrt{\bar{W}(\bm{m},N,\bar{p})},
\qquad \bm{x},\bm{m}\in\mathcal{X},
\label{ortphiderf}\\
\bar{W}(\bm{m},N,\bar{p})&\eqdef \binom{N}{\bm{m}}(\bar{\bm{p}})^{\bm{m}},\qquad
\sum_{\bm{m}\in\mathcal{X}}\bar{W}(\bm{m},N,\bar{p})=\Bigl(1+\sum_{j=1}^n\bar{p}_j\Bigr)^N.
\label{baretaW}
\end{align}

\subsection{Exactly solvable multi-Krawtchouk fermion}
\label{sec:mKraf}

Spinless fermions $\{c_{\bm x}\}$, $\{c_{\bm x}^\dagger\}$ defined on the integer lattice $\cX$ 
obey the canonical anti-commutation relations
\begin{equation}
\{c_{\bm x},c_{\bm y}^\dagger\}=\delta_{\bm{x},\bm{y}},\quad \{c_{\bm x},c_{\bm y}\}=0,\quad
\{c_{\bm x}^\dagger,c_{\bm y}^\dagger\}=0, \qquad \bm{x}, \bm{y}\in\cX.
\label{acom}
\end{equation}
Corresponding to the exactly solvable real symmetric matrix $\mathcal{H}$ \eqref{Hdef},
exactly solvable fermion Hamiltonian $\mathcal{H}_f$ is introduced \cite{solvfermi, widefermi},
\begin{equation}
\mathcal{H}_f\eqdef \sum_{\bm{x},\bm{y}\in\cX}c_{\bm x}^\dagger\mathcal{H}(\bm{x},\bm{y})c_{\bm y}.
\label{Hfdef}
\end{equation}
By introducing the momentum space fermion operators 
$\{\hat{c}_{\bm m}\}$, $\{\hat{c}_{\bm m}^\dagger\}$, $\bm{m}\in\mathcal{X}$,
\begin{align}
\hat{c}_{\bm m}&\eqdef\sum_{\bm{x}\in\mathcal{X}}\hat{\phi}_{\bm m}(\bm{x})c_{\bm x},\ \
\hat{c}_{\bm m}^\dagger
=\sum_{\bm{x}\in\mathcal{X}}\hat{\phi}_{\bm m}(\bm{x})c_{\bm x}^\dagger\ \Leftrightarrow \
{c}_{\bm x}=\sum_{{\bm m}\in\mathcal{X}}\hat{\phi}_{\bm m}(\bm{x})\hat{c}_{\bm m},\ \
{c}_{\bm x}^\dagger=\sum_{{\bm m}\in\mathcal{X}}\hat{\phi}_{\bm m}(\bm{x})\hat{c}_{\bm m}^\dagger,
\label{momrep}\\
&\hspace{2cm} \Longrightarrow \ \{\hat{c}_{\bm m}^\dagger,\hat{c}_{\bm{m}'}\}
=\delta_{{\bm m}\,{\bm m}'},\quad  
\{\hat{c}_{\bm m}^\dagger,\hat{c}_{\bm{m}'}^\dagger\}=0=\{\hat{c}_{\bm m},\hat{c}_{\bm{m}'}\},
\label{chatcom0}
\end{align}
the fermion Hamiltonian $\mathcal{H}_f$ is diagonalised
\begin{align}
 &\mathcal{H}_f=\sum_{{\bm m},{\bm m}',{\bm x},{\bm y}\in\mathcal{X}}
\hat{\phi}_{\bm m}({\bm x})
\mathcal{H}({\bm x},{\bm y})\hat{\phi}_{\bm{m}'}(\bm{y})\hat{c}_{\bm m}^\dagger\hat{c}_{\bm{m}'}
=\sum_{{\bm m},{\bm m}',{\bm x}\in\mathcal{X}}
\mathcal{E}({\bm m}')\hat{\phi}_{\bm m}({\bm x})
\hat{\phi}_{\bm{m}'}({\bm x})\hat{c}_{\bm m}^\dagger\hat{c}_{\bm{m}'}\n
&\phantom{\mathcal{H}_f}=\sum_{{\bm m}\in\mathcal{X}}
\mathcal{E}({\bm m})\hat{c}_{\bm m}^\dagger\hat{c}_{\bm m},
\label{Hjdiag}\\
&\hspace{2cm} \Longrightarrow \ [\mathcal{H}_{f},\hat{c}_{\bm m}^\dagger]
=\mathcal{E}({\bm m})\hat{c}_{\bm m}^\dagger,
\qquad [\mathcal{H}_{f},\hat{c}_{\bm m}]=-\mathcal{E}({\bm m})\hat{c}_{\bm m}.
\label{Hfcom10}
\end{align}
This fermion system has nearest neighbour interactions as is clear from the form of $\mathcal{H}$ \eqref{Hdef}.

\section{Multi-Meixner fermions }
\label{sec:Meif}

Multivariate Meixner polynomials, as hypergeometric orthogonal polynomials with the
negative multinomial distributions, have been discussed by many authors \cite{Gri12, I11}.
The main results of multivariate Meixner  polynomials obtained in \cite{mKra} are summarised as follows.

\subsection{Multivariate Meixner polynomials}
\label{sec:mMeipoly}

The  multivariate Meixner polynomials $\{P_{\bm m}({\bm x})\}$ are 
 defined by a positive constant $\beta>0$, an integer $n\ge2$  and $n$  distinct positive numbers
$c_i>0$,  $i=1,\ldots, n$, 
on a semi-infinite $n$-dimensional integer lattice $\mathcal{X}$,
\begin{equation}
 \bm{c}=(c_1,\ldots,c_n)\in\mathbb{R}_{>0}^n,\quad  |c|\eqdef \sum_{i=1}^nc_i,\quad
 \mathcal{X}=\mathbb{N}_0^n.
 \label{XMdef}
\end{equation}
The multivariate Meixner $\{P_{\bm m}({\bm x})\}$ are orthogonal with respect to the 
{\em negative multinomial distribution} \cite{Gri12}
\begin{align}
& \sum_{{\bm x}\in\cX}W(\bm{x},\beta,\bm{c})P_{\bm m}({\bm x})P_{\bm {m}'}({\bm x})=0,
\quad \bm{m}\neq\bm{m}'\in\cX,
\label{pMortho}\\
&\qquad W(\bm{x},\beta,\bm{c})\eqdef\frac{(\beta)_{|x|}\bm{c}^{\bm x}}{\bm{x}!}(1-|c|)^{\beta}.
\label{WMei}
\end{align}
The summability of $W(\bm{x},\beta,\bm{c})$ requires $|c|<1$.
They form a complete set of eigenvectors of a real symmetric $|\cX|\times|\cX|$ matrix $\mathcal{H}$,
\begin{align}
&B_i(\bm{x})\eqdef \beta+|x|,\quad  D_i(\bm{x})\eqdef c_i^{-1}x_i,\qquad  i=1,\ldots,n,
\label{BDMeidef}\\
\mathcal{H}(\bm{x},\bm{y})&\eqdef\sum_{j=1}^n
\left[\bigl(B_j(\bm{x})+D_j(\bm{x})\bigr)\,\delta_{\bm{x}\,\bm{y}}
-\sqrt{B_j(\bm{x})D_j(\bm{x}+\bm{e}_j)}\,\delta_{\bm{x}+\bm{e}_j\,\bm{y}}\right.\n
&\left.\hspace{4cm}
-\sqrt{B_j(\bm{x}-\bm{e}_j)D_j(\bm{x})}\,\delta_{\bm{x}-\bm{e}_j\,\bm{y}}\right],
\qquad \quad\, \bm{x},\bm{y}\in\cX,
\label{HdefM}\\
&\sum_{\bm{y}\in\cX}\mathcal{H}(\bm{x},\bm{y})\sqrt{W(\bm{y},\beta,\bm{c})}\,P_{\bm m}(\bm{y})
=\mathcal{E}(\bm{m})\sqrt{W(\bm{x},\beta,\bm{c})}\,P_{\bm m}(\bm{x}),\quad \bm{m}\in\cX.
\label{HPeigM}
\end{align}
The eigenvalue $\mathcal{E}(\bm{m})$ has a linear spectrum
\begin{equation}
\mathcal{E}(\bm{m})\eqdef\sum_{j=1}^nm_j\lambda_j\ge0,\qquad \bm{m}\in\cX,
\tag{I.4.14}
\end{equation}
in which $\lambda_j>0$ is the $j$-th root of a degree $n$ characteristic polynomial $\mathcal{F}(\lambda)$
of an $n\times n$ positive definite symmetric matrix $F(\bm{c})$  depending on $\{c_i\}$,
\begin{equation}
0=\mathcal{F}(\lambda)
\eqdef Det\bigl(\lambda I_n-F(\bm{c})\bigr),\quad F(\bm{c})_{i\,j}\eqdef -1+c_i^{-1}\delta_{i\,j}.
\tag{I.4.15}
\end{equation}
The multivariate Meixner polynomial  $P_{\bm{m}}(\bm{x})$ is a terminating   $(n+1,2n+2)$ hypergeometric function 
of Aomoto-Gelfand  {\rm \cite{AK,gelfand,mizu}}
\begin{gather}
\label{PmM}
P_{\bm{m}}(\bm{x};\beta,{\bm u})
\eqdef \sum_{\substack{\sum_{i,j}c_{ij}\\
(c_{ij})\in {\mathbb M}_{n}({\mathbb N}_{0})}}
\frac{\prod\limits_{i=1}^{n}(-x_{i})_{\sum\limits_{j=1}^{n}c_{ij}}
\prod\limits_{j=1}^{n}(-m_{j})_{\sum\limits_{i=1}^{n}c_{ij}}}
{(\beta)_{\sum_{i,j}c_{ij}}} \; \frac{\prod(u_{ij})^{c_{ij}}}{\prod c_{ij}!}.
\end{gather}
The $n\times n$ matrix $u_{i\,j}$  is defined by
\begin{equation}
u_{i\,j}\eqdef\frac{\lambda_j}{\lambda_j-c_i^{-1}},\quad i,j=1,\ldots,n.
\tag{I.4.17}
\end{equation}
The orthogonality relation of the multivariate Meixner polynomials \eqref{pMortho}  now reads
\begin{align}
&\sum_{\bm{x}\in\mathcal{X}}W(\bm{x},\beta,\bm{c})
P_{\bm{m}}(\bm{x};\beta,{\bm u})P_{\bm{m}'}(\bm{x};\beta,{\bm u})
=\frac{1}{\bar{W}({\bm m},\beta,\bar{\bm{c}})}\,\delta_{{\bm m}\,{\bm m}'},
\quad {\bm m},{\bm m}'\in\cX,
\label{ortrelM}\\
&\hspace{3cm}\bar{W}({\bm m},\beta,\bar{\bm{c}})
\eqdef \frac{(\beta)_{|m|}\bar{\bm c}^{\bm m}}{{\bm m}!},
\quad \sum_{\bm{m}\in{\mathcal X}}\bar{W}({\bm m},\beta,\bar{\bm{c}})=(1-|\bar{c}|)^{-\beta},
\label{barWdef}\\
& \hspace{4cm} \bar{c}_j\eqdef \frac{1-|c|}{1-|c|+\sum_{i=1}^nc_iu_{i\,j}^2},\quad j=1,\ldots,n,
\tag{I.4.20}
\end{align}
leading to the complete set of orthonormal eigenvectors $\{\hat{\phi}_{\bm m}(\bm{x})\}$,
\begin{align}
&\quad \sum_{\bm{y}\in\cX}\mathcal{H}(\bm{x},\bm{y})\hat{\phi}_{\bm{m}}(\bm{y};\beta,{\bm u})
=\mathcal{E}(\bm{m})\hat{\phi}_{\bm{m}}(\bm{x};\beta,{\bm u}),\quad \bm{x},\bm{m}\in\cX,
\label{MHeig}\\
&\quad \sum_{\bm{x}\in{\mathcal X}}\hat{\phi}_{\bm{m}}(\bm{x};\beta,{\bm u})
\hat{\phi}_{\bm{m}'}(\bm{x};\beta,{\bm u})
=\delta_{\bm{m},\bm{m}'}, \qquad \quad \bm{m},\bm{m}'\in{\mathcal  X},
\label{Mortrel2}\\
&\quad \sum_{\bm{m}\in{\mathcal X}}\hat{\phi}_{\bm{m}}(\bm{x};\beta,{\bm u})
\hat{\phi}_{\bm{m}}(\bm{y};\beta,{\bm u})
=\delta_{\bm{x},\bm{y}}, \qquad \qquad \bm{x},\bm{y}\in{\mathcal  X},
\label{Mortrel3}\\
\hat{\phi}_{\bm{m}}(\bm{x};\beta,{\bm u})&
\eqdef \sqrt{W(\bm{x},\beta,\bm{c})}
P_{\bm{m}}(\bm{x};\beta,{\bm u})\sqrt{\bar{W}(\bm{m},\beta,\bar{\bm{c}})},
\qquad \bm{x},\bm{m}\in{\mathcal X}.
\label{ortphiderfM}
\end{align}

\subsection{Exactly solvable multi-Meixner fermion}
\label{sec:mMeif}

Corresponding to the exactly solvable real symmetric matrix $\mathcal{H}$ \eqref{HdefM},
exactly solvable fermion Hamiltonian $\mathcal{H}_f$ is introduced \cite{solvfermi, widefermi},
\begin{equation}
\mathcal{H}_f\eqdef \sum_{\bm{x},\bm{y}\in\cX}c_{\bm x}^\dagger\mathcal{H}(\bm{x},\bm{y})c_{\bm y},
\label{HfdefM}
\end{equation}
in which, as before, spinless fermions $\{c_{\bm x}\}$, $\{c_{\bm x}^\dagger\}$ defined on the integer lattice $\cX$ 
obey the canonical anti-commutation relations \eqref{acom}.

By introducing the momentum space fermion operators 
$\{\hat{c}_{\bm m}\}$, $\{\hat{c}_{\bm m}^\dagger\}$, $\bm{m}\in\mathcal{X}$,
\begin{align}
\hat{c}_{\bm m}&\eqdef\sum_{\bm{x}\in\mathcal{X}}\hat{\phi}_{\bm m}(\bm{x};\beta,{\bm u})c_{\bm x},\ \
\hat{c}_{\bm m}^\dagger
=\sum_{\bm{x}\in\mathcal{X}}\hat{\phi}_{\bm m}(\bm{x};\beta,{\bm u})c_{\bm x}^\dagger
\label{momrepM1}\\
\Longleftrightarrow \
{c}_{\bm x}&=\sum_{{\bm m}\in\mathcal{X}}\hat{\phi}_{\bm m}(\bm{x};\beta,{\bm u})\hat{c}_{\bm m},\ \
{c}_{\bm x}^\dagger
=\sum_{{\bm m}\in\mathcal{X}}\hat{\phi}_{\bm m}(\bm{x};\beta,{\bm u})\hat{c}_{\bm m}^\dagger,
\label{momrepM2}\\
&\hspace{2cm} \Longrightarrow \ \{\hat{c}_{\bm m}^\dagger,\hat{c}_{\bm{m}'}\}
=\delta_{{\bm m}\,{\bm m}'},\quad  
\{\hat{c}_{\bm m}^\dagger,\hat{c}_{\bm{m}'}^\dagger\}=0=\{\hat{c}_{\bm m},\hat{c}_{\bm{m}'}\},
\label{chatcom0M}
\end{align}
the fermion Hamiltonian $\mathcal{H}_f$ is diagonalised
\begin{align}
 \mathcal{H}_f&=\sum_{{\bm m},{\bm m}',{\bm x},{\bm y}\in\mathcal{X}}
\hat{\phi}_{\bm m}(\bm{x};\beta,{\bm u})\mathcal{H}({\bm x},{\bm y})
\hat{\phi}_{\bm{m}'}(\bm{y};\beta,{\bm u})\hat{c}_{\bm m}^\dagger\hat{c}_{\bm{m}'}\n
&=\sum_{{\bm m},{\bm m}',{\bm x}\in\mathcal{X}}\mathcal{E}({\bm m}')
\hat{\phi}_{\bm m}(\bm{x};\beta,{\bm u})
\hat{\phi}_{\bm{m}'}(\bm{x};\beta,{\bm u})\hat{c}_{\bm m}^\dagger\hat{c}_{\bm{m}'}\n
&=\sum_{{\bm m}\in\mathcal{X}}\mathcal{E}({\bm m})\hat{c}_{\bm m}^\dagger\hat{c}_{\bm m},
\label{HjdiagM}\\
&\hspace{2cm} \Longrightarrow \ [\mathcal{H}_{f},\hat{c}_{\bm m}^\dagger]
=\mathcal{E}({\bm m})\hat{c}_{\bm m}^\dagger,
\qquad [\mathcal{H}_{f},\hat{c}_{\bm m}]=-\mathcal{E}({\bm m})\hat{c}_{\bm m}.
\label{Hfcom10M}
\end{align}
This fermion system has nearest neighbour interactions as is clear from the form of $\mathcal{H}$ \eqref{HdefM}.

\section{Rahman like fermions}
\label{sec:Rah}
 Rahman  polynomials have long been investigated as  typical examples of multivariate orthogonal polynomials,
\cite{gr1}--\cite{gr3}, 
\cite{coo-hoa-rah77, HR0, HR,IT, I}.
Two types of  Rahman like polynomials introduced here are constructed as eigenvectors of certain
{\em reversible} Markov chain matrices $\mathcal{K}^{(i)}$, $i=1,2$  \cite{Rah}, which will be cited as II.
This method is very different from those of the existing theories and models. 
The main results of Rahman  like polynomials \cite{Rah} of type (1) and type (2) are summarised as follows.
Both types of   Rahman like polynomials $\{P_{\bm m}({\bm x})\}$ 
 depend on  two positive integers $N$ and $n$ ($N>n\ge2$) and $2n$  distinct positive numbers
$0<\alpha_i,\beta_i<1$,  $i=1,\ldots, n$, with $|\beta|<1$,
on a finite $n$-dimensional integer lattice $\mathcal{X}$,
\begin{equation}
 \mathcal{X}=\{\bm{x}\in\mathbb{N}_0^n\ |\, |x|\le N\},\quad \bm{\alpha}=(\alpha_1,\ldots,\alpha_n),\quad
 \bm{\beta}=(\beta_1,\ldots,\beta_n),\ |\beta|=\sum_{i=1}^n\beta_i.
 \label{XRdef}
\end{equation}
These are the parameters of the binomial ($\bm{\alpha}$) and multinomial ($\bm{\beta}$) distributions,
\begin{align}
&W_1(x,y,\alpha)\eqdef \binom{y}{x}\alpha^x(1-\alpha)^{y-x}>0,
\quad \sum_{x=0}^yW_1(x,y,\alpha)=1,
\label{W1def}\\
&W_n({\bm x},N,{\bm \beta})\eqdef 
\frac{N!\cdot(1-|\beta|)^{N-|x|}}{x_1!\cdots x_n!(N-|x|)!}\cdot\prod_{i=1}^n\beta_i^{x_i}
=\binom{N}{\bm{x}}\beta_0^{x_0}\bm{\beta}^{\bm{x}}>0,
\label{Wndef}\\
&\sum_{{\bm x}\in\mathcal{X}}W_n({\bm x}, N,{\bm \beta})=1,\quad    x_0\eqdef
N-|x|,\  \beta_0\eqdef 1-|\beta|.  
\nonumber
\end{align}
Two types of Markov chain matrices $\mathcal{K}^{(i)}$, $i=1,2$ are constructed 
by certain convolutions of the multinomial distribution and $n$-tuple of the binomial distributions,
\begin{align} 
  & \mathcal{K}^{(1)}({\bm x},{\bm y})
 \eqdef\sum_{{\bm z}\in\mathcal{X}}W_n({\bm x}-{\bm z}, N-|z|,{\bm \beta})
 \prod_{i=1}^nW_1(z_i, y_i,\alpha_i)>0,
\label{Kn1def}\\
 & \mathcal{K}^{(2)}({\bm x},{\bm y})
 \eqdef\sum_{{\bm z}\in\mathcal{X}}W_n({\bm x}-{\bm z}, N-|y|,{\bm \beta})
 \prod_{i=1}^nW_1(z_i, y_i,\alpha_i)>0.
\label{Kn2def}
\end{align}
The convention is that the transition probability matrix per unit time interval 
$\mathcal{K}(\bm{x},\bm{y})$ on $\mathcal{X}$ means
the transition from an initial point $\bm{y}$ to a final point $\bm{x}$ with  
 \begin{equation}
  \mathcal{K}^{(i)}(\bm{x},\bm{y})>0,\quad\sum_{x\in{X}}\mathcal{K}^{(i)}(\bm{x},\bm{y})=1,\quad i=1,2.
  \label{basK}
\end{equation}
Positive $\mathcal{K}$  means wide range interactions as  
every pair of points $\bm{x}$ and $\bm{y}$ is connected. 

Reversibility means that $\mathcal{K}$ has a reversible distribution $W_n(\bm{x},N,\bm{\eta})$ satisfying
\begin{equation}
\mathcal{K}(\bm{x},\bm{y})W_n(\bm{y},N,\bm{\eta})=
\mathcal{K}(\bm{y},\bm{x})W_n(\bm{x},N,\bm{\eta}).
\end{equation}
This condition determines the probability parameter $\bm{\eta}$
as a function of $\bm{\alpha}$ and $\bm{\beta}$ in
type (1) and (2);
\begin{align}
\text{type (1)}:\qquad & \eta_i=\frac{\beta_i}{1-\alpha_i}\frac1{D_n},\quad i=1,\ldots,n,
\qquad D_n\eqdef 1+\sum_{k=1}^n\frac{\alpha_k\beta_k}{1-\alpha_k},
\tag{II.2.12}\\
%
\text{type (2)}: \qquad& \eta_i=\frac{\beta_i}{1-\alpha_i}\frac1{D_n},\quad i=1,\ldots,n,
\qquad D_n\eqdef 1+\sum_{k=1}^n\frac{\beta_k}{1-\alpha_k}.
\tag{II.2.17}
\end{align}

As the reversible distribution is the stationary distribution, both types of 
Rahman like polynomials are orthogonal with respect to the stationary distribution $W_n({\bm x},N,{\bm \eta})$
with each own $\bm{\eta}$,
\begin{align}
&\sum_{\bm{x}\in\cX}W_n({\bm x},N,{\bm \eta})P_{\bm m}({\bm x})P_{\bm{m}'}({\bm x})=0,
\quad \bm{m}\neq\bm{m}'\in\cX,
\label{Rorthogen}\\
&W_n({\bm x}, N,{\bm \eta})\eqdef 
\frac{N!\cdot(1-|\eta|)^{N-|x|}}{x_1!\cdots x_n!(N-|x|)!}\cdot\prod_{i=1}^n\eta_i^{x_i}
=\binom{N}{\bm{x}}\eta_0^{x_0}\bm{\eta}^{\bm{x}}>0,
\label{Wndefgen}\\
&\sum_{{\bm x}\in\mathcal{X}}W_n({\bm x},N,{\bm \eta})=1,\quad    x_0\eqdef
N-|x|,\  \eta_0\eqdef 1-|\eta|.
\nonumber
\end{align}

In the following, the basic properties of Rahman like polynomials of type (1) and (2) are summarised in 
\S\ref{sec:Rahpoly1} and \S\ref{sec:Rahpoly2} and the corresponding exactly solvable fermion systems
are presented at the end of each subsection.
For the details of the derivation of Rahman like polynomials, consult \cite{Rah}.

\subsection{Rahman like polynomials type (1)}
\label{sec:Rahpoly1}
Rahman like polynomials of type (1) are {\em left eigenvectors of reversible} Markov chain matrix
$\mathcal{K}^{(1)}$ with a multiplicative spectrum,
\begin{align}
&\sum_{{\bm x}\in\mathcal{X}}\mathcal{K}^{(1)}({\bm x},{\bm y})
P_{\bm{m}}(\bm{x};{\bm u})=\mathcal{E}({\bm m})P_{\bm{m}}(\bm{y};{\bm u}),
\quad \forall{\bm m}\in\mathcal{X},
\label{KnPmeig}\\
&\hspace{4cm}\mathcal{E}({\bm m})=\prod_{i=1}^n\lambda_i^{m_i},
\tag{II.3.25}
\end{align}
in which $\{\lambda_i\}$ $i=1,\ldots,n$ are 
the  roots   of the characteristic equation,
\begin{equation}
Det\left(\lambda\,I_n-F^{(1)}({\bm \alpha},{\bm \beta})\right)=0,\quad 
F^{(1)}({\bm \alpha},{\bm \beta})_{i\,j}\eqdef-\alpha_i\beta_j+\alpha_i\delta_{i\,j},\quad i,j=1,\ldots,n.
\tag{II.3.7}
\end{equation}
It should be stressed that $-1<\mathcal{E}(\bm{m})\le1$ due to Perron-Frobenius theorem of positive matrices.
This is in good contrast with the multivariate Krawtchouk I(3.13)
and Meixner I(4.14) polynomials.
The Rahman like polynomials of type (1) $P_{\bm{m}}(\bm{x};{\bm u})$ 
are $(n+1,2n+2)$ type terminating hypergeometric function of 
Aomoto-Gelfand {\rm \cite{AK,gelfand, mizu}} depending on a set of parameters 
$\bm{u}=(u_{i,j})$, $i,j=1,\ldots,n$, 
\begin{equation}
u_{i,j}=\frac{\alpha_i(\lambda_j-1)}{\lambda_j-\alpha_i},
\tag{II.3.16}
\end{equation}
\begin{gather}
\label{PmR1}
P_{\bm{m}}(\bm{x};{\bm u})
\eqdef \sum_{\substack{\sum_{i,j}c_{ij}\leq N\\
(c_{ij})\in M_{n}({\mathbb N_{0}})}}
\frac{\prod\limits_{i=1}^{n}(-x_{i})_{\sum\limits_{j=1}^{n}c_{ij}}
\prod\limits_{j=1}^{n}(-m_{j})_{\sum\limits_{i=1}^{n}c_{ij}}}
{(-N)_{\sum_{i,j}c_{ij}}} \; \frac{\prod(u_{ij})^{c_{ij}}}{\prod c_{ij}!}.
\end{gather}
The above general form is the consequence of the orthogonality \eqref{Rorthogen} \cite{mizu} and 
the eigenvalues $\{\lambda_i\}$ II(3.7) 
and the explicit
form of $u_{i,j}$ II(3.16)
are determined by the degree one solutions of \eqref{KnPmeig}.

Since the distribution $W_n(\bm{x}, N,\bm{\eta})$ with the probability $\bm{\eta}$ in II(2.12)
 is the {\em reversible distribution} of $\mathcal{K}^{(1)}(\bm{x},\bm{y})$,
the real symmetric matrix (Hamiltonian) is obtained by
\begin{equation}
\mathcal{H}^{(1)}(\bm{x},\bm{y})\eqdef\frac1{\sqrt{W_n(\bm{x}, N,\bm{\eta})}}\,
\mathcal{K}^{(1)}(\bm{x},\bm{y})
\,\sqrt{W_n(\bm{y}, N,\bm{\eta})},\quad \bm{x},\bm{y}\in\cX.
\label{HR1def}
\end{equation}
With the explicit expression of the type (1) Rahman like polynomials \eqref{PmR1}, 
the orthogonality relation \eqref{Rorthogen}
 now reads
\begin{align}
\sum_{\bm{x}\in\mathcal{X}}W_n(\bm{x}, N,\bm{\eta})P_{\bm{m}}(\bm{x};\bm{u})P_{\bm{m}'}(\bm{x};\bm{u})
&=\frac{\delta_{\bm{m}\,\bm{m}'}}{\binom{N}{\bm{m}}(\bar{\bm{p}})^{\bm{m}}},\qquad 
(\bar{\bm{p}})^{\bm{m}}\eqdef\prod_{j=1}^n\bar{p}_j^{m_j},
\label{R1orth}\\
\bar{p}_j&=\Bigl(\sum_{i=1}^n\eta_iu_{i,j}^2-1\Bigr)^{-1}>0,\quad j=1,\ldots,n,
\label{R1pddef}
\end{align}
leading to the complete set of orthonormal eigenvectors $\{\hat{\phi}_{\bm m}(\bm{x})\}$,
\begin{align}
&\sum_{\bm{y}\in\mathcal{X}}\mathcal{H}^{(1)}(\bm{x},\bm{y})\hat{\phi}_{\bm{m}}(\bm{y};\bm{u})
=\mathcal{E}(\bm{m})\hat{\phi}_{\bm{m}}(\bm{x};\bm{u}),\quad \bm{x},\bm{m}\in\cX,
\label{R1Hphin}\\
&\sum_{\bm{x}\in\mathcal{X}}\hat{\phi}_{\bm{m}}(\bm{x};\bm{u})\hat{\phi}_{\bm{m}'}(\bm{x};\bm{u})
=\delta_{\bm{m},\bm{m}'}, \n
&\sum_{\bm{m}\in\mathcal{X}}\hat{\phi}_{\bm{m}}(\bm{x};\bm{u})\hat{\phi}_{\bm{m}}(\bm{y};\bm{u})
=\delta_{\bm{x},\bm{y}}, \qquad \qquad \bm{x},\bm{y}, \bm{m},\bm{m}'\in\mathcal{X},
\label{R1ortrel}\\
\hat{\phi}_{\bm{m}}(\bm{x};\bm{u})&\eqdef
\sqrt{W_n(\bm{x}, N,\bm{\eta})}P_{\bm{m}}(\bm{x};\bm{u})\sqrt{\bar{W}_n(\bm{m}, N,\bar{\bm p})},
\qquad \bm{x},\bm{m}\in\mathcal{X},
\label{R1ortphiderf}\\
\bar{W}_n(\bm{m}, N,\bar{\bm p})&\eqdef \binom{N}{\bm{m}}(\bar{\bm{p}})^{\bm{m}},\qquad
\sum_{\bm{m}\in\mathcal{X}}\bar{W}_n(\bm{m}, N,\bar{\bm p})=\Bigl(1+\sum_{j=1}^n\bar{p}_j\Bigr)^N.
\label{R1baretaW}
\end{align}

\paragraph{Exactly solvable type (1) Rahman like fermion}

Corresponding to the exactly solvable real symmetric matrix $\mathcal{H}^{(1)}$ \eqref{HR1def},
exactly solvable fermion Hamiltonian $\mathcal{H}_f^{(1)}$ is introduced
\begin{equation}
\mathcal{H}^{(1)}_f\eqdef 
\sum_{\bm{x},\bm{y}\in\cX}c_{\bm x}^\dagger\mathcal{H}^{(1)}(\bm{x},\bm{y})c_{\bm y},
\label{HR1fdef}
\end{equation}
in which, as before, spinless fermions 
$\{c_{\bm x}\}$, $\{c_{\bm x}^\dagger\}$ defined on the integer lattice $\cX$ 
obey the canonical anti-commutation relations \eqref{acom}. 
The diagonalisation of $\mathcal{H}^{(1)}_f$ \eqref{HR1fdef}
goes exactly the same as that of multivariate 
Krawtchouk \S\ref{sec:mKraf} and multivariate Meixner 
\S\ref{sec:mMeif} fermion's cases \cite{solvfermi, widefermi}.
\subsection{Rahman like polynomials type (2)}
\label{sec:Rahpoly2}

The formulas for type (2) polynomials look very similar to those of type (1).
The energy spectrum $\mathcal{E}(\bm{m})$ are also multiplicative II(3.25)
and the eigenvalues $\{\lambda_i\}$, $i=1,\ldots,n$ are the roots 
 of the characteristic equation
\begin{equation}
Det\left(\lambda\,I_n-F^{(2)}({\bm \alpha},{\bm \beta})\right)=0,\quad 
F^{(2)}({\bm \alpha},{\bm \beta})_{i\,j}\eqdef-\beta_j+\alpha_i\delta_{i\,j},\quad i,j=1,\ldots,n,
\tag{II.3.12}
\end{equation}
and the parameters $\{u_{i,j}\}$ are 
\begin{equation}
u_{i,j}=\frac{\lambda_j-1}{\lambda_j-\alpha_i}.
\tag{II.3.17}
\end{equation}
The polynomial $P_{\bm m}(\bm{x};\bm{u})$ has the same expression as type (1) \eqref{PmR1}.
The other formulas from \eqref{HR1def} to \eqref{HR1fdef} need be changed (1) to (2).
So they are not repeated here.

\section*{Declarations}
\begin{itemize}
\item Funding: No funds, grants, or other support was received.
\item Data availability statement: Data sharing not applicable to this article as 
no datasets were generated or analysed during the current study.
\item Competing Interests: The author has no competing interests to declare that 
are relevant to the content of this article.
\end{itemize}



\end{document}